# THUIR2 at NTCIR-16 Session Search (SS) Task

Weihang Su[2], Xiangsheng Li[1], Yiqun Liu[1★], Min Zhang[1], Shaoping Ma[1]
1 Department of Computer Science and Technology, Institute for Artificial Intelligence, Beijing National Research Center for Information Science and Technology, Tsinghua University, Beijing, China
2 School of Computer Science, Beijing University of Posts and Telecommunications, Beijing, China
yiqunliu@tsinghua.edu.cn

## ABSTRACT

Our team(THUIR2) participated in both FOSS and POSS subtasks of the NTCIR-16[1] Session Search (SS) Task[2]. This paper describes our approaches and results. In the FOSS subtask, we submit five runs by using a learning-to-rank model and a fine-tuned pre-trained language model. We fine-tune the pre-trained language model with both ad-hoc data and session information and then assembled them by a learning-to-rank method. The assembled model achieves the best performance among all participants in the preliminary evaluation. In the POSS subtask, we used an assembled model which also achieves the best performance in the preliminary evaluation.

## KEYWORDS

Session Search, Learning-to-rank, Pre-trained Language Model

## TEAM NAME

THUIR2

## SUBTASKS

Session Search (FOSS, POSS)

## 1 INTRODUCTION

In NTCIR-16, our team participated in the Session Search task, including both FOSS and POSS subtasks. In recent years, pre-trained language model fine-tuned on downstream tasks have achieved state-of-the-art performance of document ranking tasks. These models estimate the relevance between queries and documents based on contextualized matching signals. "Session search task aims to provide an optimized document ranking list by utilizing user interaction behaviors within a search session"[1]. We thus apply users' behavior information during the fine-tuning stage to verify the effects of users' behavior information in a search session.

In the FOSS subtask, we submit five runs by using a learning-to-rank model and a fine-tuned pre-trained language model. Specifically, two runs are based on learning-to-rank model and three runs are based on fine-tuned pre-trained language models. One of the learning-to-rank models only utilizes the score of traditional IR methods like BM25, TF-IDF, and F1-EXP. The other model assembled the scores of two fine-tuned pre-trained language model as well as traditional IR methods. In the POSS subtask, we submit the result of the same assembled model which achieves the best performance among all participants in the preliminary evaluation.

The results show that the learning-to-rank models perform better than a single model. Assembled traditional methods perform better than assembled pre-trained language model in the final evaluation which is not what we expected. This unexpected result shows that traditional methods are still good solutions in document ranking tasks.

## 2 FOSS SUBTASK

In the FOSS subtask, we submit five runs which are shown in Table 1. We tried traditional methods, pre-trained language model and learning-to-rank models. To be specific, we tried traditional methods BM25, TF-IDP, and F1-EXP. We fine-tuned pre-trained language model with both ad-hoc data and session information. We used the learning-to-rank model lambdaMART to assemble the scores of those methods.

The details of our runs are described in this section.

### 2.1 The Framework of Assembled Models

*2.1.1 Assembled Traditional Methods(ATM) Model.* We design the Assembled Traditional Methods(ATM) model to rerank the candidate documents via traditional methods and a learning-to-rank model. The framework of the ATM model is shown in Figure 1. Firstly, we filter the candidate documents by length, keyword, and key phrased which is detailed in Section 2.2. Then we calculated the scores of each query-document pair via several traditional methods and used a learning-to-rank model to assemble them. Finally, we rank the candidate documents by the score of the learning-to-rank model which is detailed in Section 2.3.

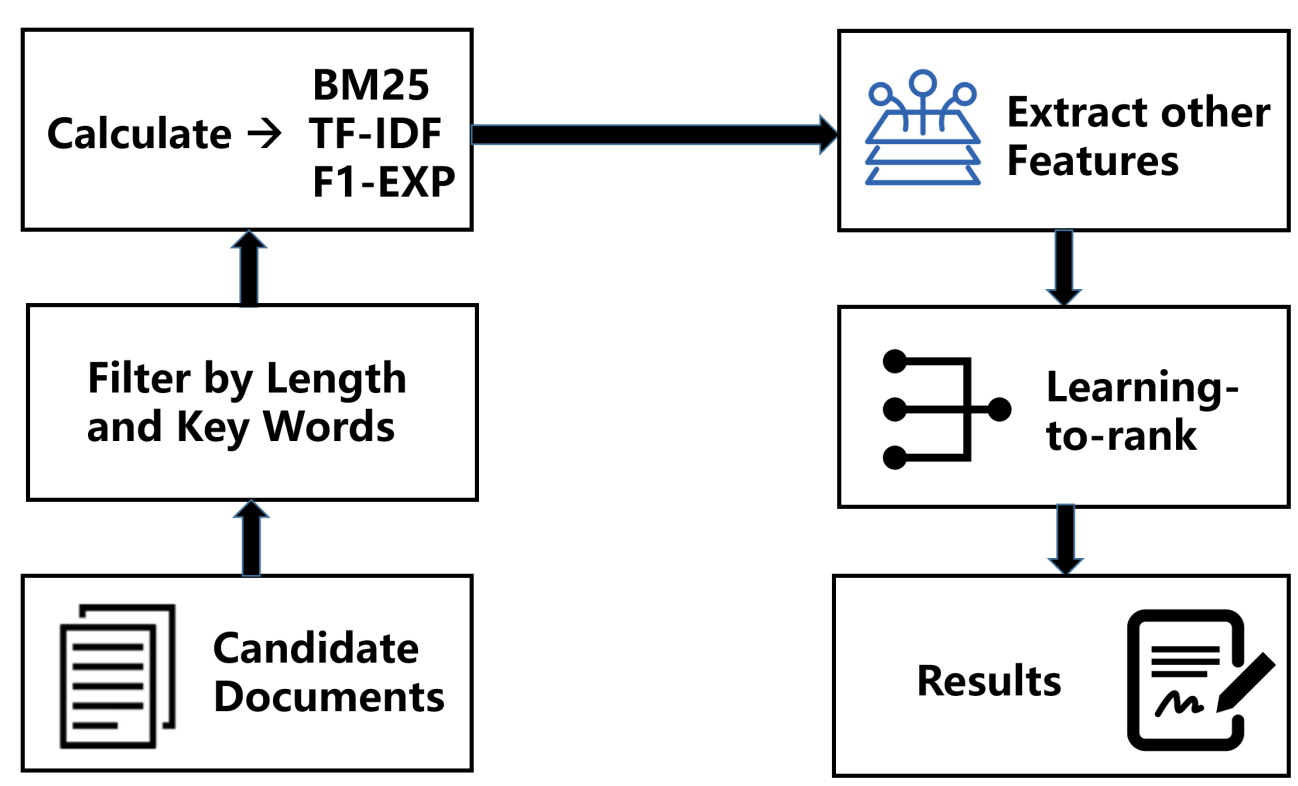

**Figure 1: The Framework of ATM Model**

*2.1.2 Assembled Pre-trained Models and Traditional Methods (PMTM) Model.* In this method, we fine-tuned the pre-trained language model BERT with both ad-hoc data and users' behavior information(session information) which are described in detail in Section 2.4. Then we add the scores of two fine-tuned models as new features to the

[1]https://research.nii.ac.jp/ntcir/workshop/OnlineProceedings15/NTCIR/toc_ntcir.html

**Table 1: Preliminary Evaluation of Our Runs in FOSS Subtask**

| Run Name | Description | NDCG@3 | Rank |
|---|---|---|---|
| THUIR2-FOSS-NEW-2 | BM25 + TF-IDF + F1-EXP | 0.022508 | 7 |
| THUIR2-FOSS-NEW-3 | Bert with Ad-hoc Data Fine-tune | 0.013015 | 14 |
| THUIR2-FOSS-NEW-4 | Bert with Click Model Fine-tune | – | – |
| THUIR2-FOSS-NEW-5 | Three Traditional Methods and two Pre-train Models | 0.095888 | 1 |
| THUIR2-FOSS-NEW-6 | Bert with Session Data Fine-tune | 0.016912 | 10 |

ATM model. Finally, we rank the candidate documents by the score of the learning-to-rank model which is detailed in Section 2.4.3.

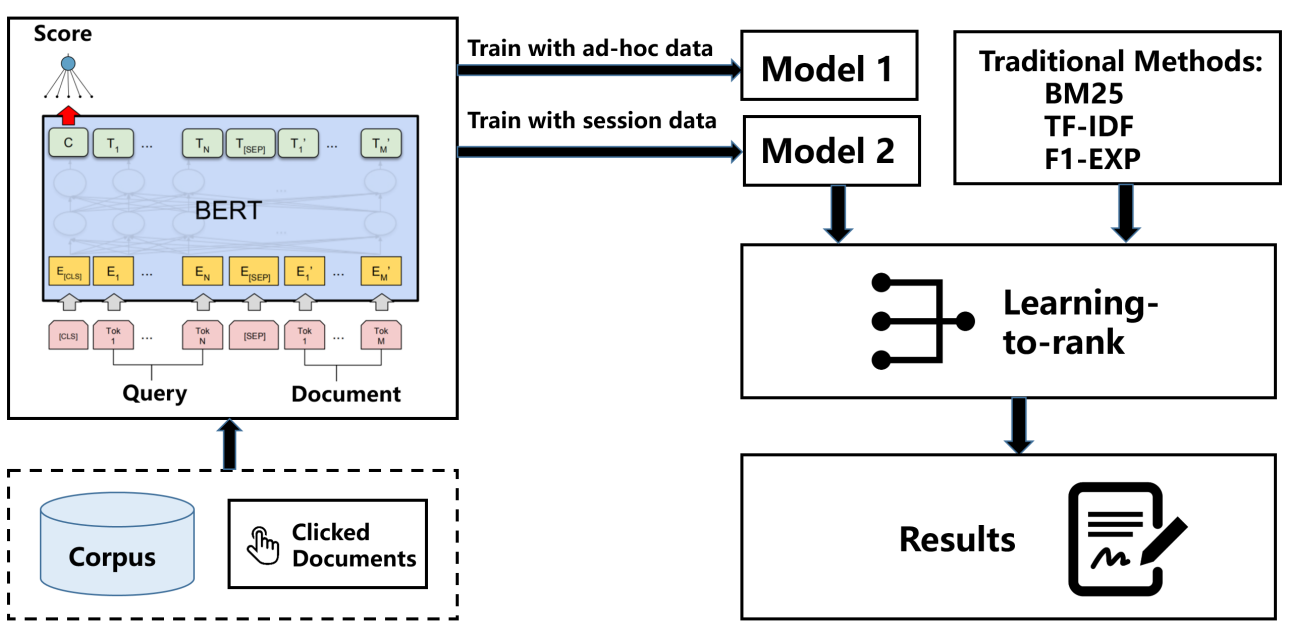

**Figure 2: The Framework of PMTM Model**

## 2.2 Data Preprocess

We filtered the candidate documents before reranking. The documents in the corpus are extracted from web pages, some of them are autogenerated meaningless websites which is shown in Figure 4. We filtered those documents by length, keywords, and key phrases. Some documents include phrases like "404 The requested resource is not found", "404 Not Found" etc are unlikely to meet users' information needs. Thus, we remove those documents before reranking. The examples of filtered out long documents are shown in Figure 3.

## 2.3 Traditional Methods

We first tried three traditional Information Retrieval(IR) methods: BM25[7], TF-IDF, and F1EXP[5]. The calculation formula for the BM25 and F1EXP scores are shown in Equation 1 and Equation 2:

$$BM25(d,q) = \sum_{i=1}^{m} \frac{IDF(ti) * TF(t_i,d) * (k_1+1)}{TF(t_i,d) + k_1 * (1 - b + b * \frac{len(d)}{avgdl})} \quad (1)$$

$$F1EXP(d,q) = \sum_{t \in q \cup d} C(t,q) * F(C(t,d)) * LN(d) * (\frac{N+1}{df(t)})^k \quad (2)$$

where $F(x) = 1 + ln(1 + ln(x)), LN(x) = \frac{avdl+s}{avdl+x*s}, C(t,q)$ is the term frequency of t in the query, $C(t,d)$ is the term frequency of t in the document and s is an adjustable parameter. We set the parameter $k1 = 2, k2 = 1, b = 0.5$ in BM25 calculation formula.

After calculating the score of each query-document pair, we extracted more features such as the length of the query, the length of the document, and the document's rank sorted by the scores of traditional IR methods. All the features are shown in Table 2. After calculating all those features, we feed them into a learning-to-rank model from Ranklib[8]. The Ranklib package provides several learning-to-rank algorithms including LambdaMART. We used the LambdaMART model with the training metric NDCG@10.

**Table 2: Features of ATM Model**

| | Features |
|---|---|
| 1 | Score of BM25 |
| 2 | Rank sorted by BM25 |
| 3 | Score of TF-IDF |
| 4 | Rank sorted by TF-IDF |
| 5 | Score of F1-EXP |
| 6 | Rank sorted by F1-EXP |
| 7 | Document Length |
| 8 | Query Length |

To our surprise, this method gets the third-highest score in the final evaluation. It's better than many runs which use large-scale pre-trained language model such as BERT and Hierarchical Behavior Aware Transformers. We guess the reason is that the final evaluation results are based on manually labeled data. When labeling, it is mainly judged according to the keywords in the query and its appearance in the document, which is very similar to the traditional method.

## 2.4 Pre-trained Language Model

Pre-trained language models such as BERT[4], has been widely used in document ranking tasks in recent years and it significantly outperforms traditional IR methods as well as other neural ranking models[3]. Thus, we tried BERT for Session Search Task and fine-tune the BERT model on ad-hoc data and session data, respectively.

*2.4.1 Fine-tune with Ad-hoc Data.* The format of training data is shown in Figure 5. For each query, the Sougou search engine show 10 pages to the user and record the clicked pages. In ad-hoc search scenarios, the clicked page is more likely to match the user's information needs. We define the clicked web page as $d^+$ and the others as $d^-$.

We use BERT as a ranking model. The score of a query document pair is defined as follows,

$$Input := [CLS]query[SEP]doc[SEP]$$

```
d366628.txt 非常抱歉 您访问的页面不存在 视频 首页 电视剧   电影 体育 综艺 动漫 频道 电视剧   偶像 爱情 军旅 历史 武侠 神话 电  影  动作 喜剧 爱情 恐怖 犯罪 剧情 动  漫  热血 青春 神魔 校园 亲子 搞笑
d363950.txt 错误 你访问的页面丢失了 页头 导航 爱奇艺视频   爱奇艺   爱奇艺   返回爱奇艺首页   返回爱奇艺首页   返回爱奇艺首页   导航 导航 原创 原创 更多频道内容在这里查看  更多频道内容在这里
d368949.txt 404 -   非常抱歉，您访问的页面不存在 -   PP视频一起玩出精彩 -   原聚力视频、PPTV聚力   电影 首页 |   电视剧   电影 体育 综艺 动漫 频道 电视剧   偶像 爱情 军旅 历史 武侠 神话 电   影
d361641.txt 非常抱歉 您访问的页面不存在 视频一起玩出精彩   原聚力视频   聚力 电影 首页 电视剧   电影 体育 综艺 动漫 频道 电视剧   偶像 爱情 军旅 历史 武侠 神话 电   影   动作 喜剧 爱情 恐怖 犯罪 剧情
d369669.txt 错误 你访问的页面丢失了 导航 更多频道内容在这里查看知道了 娱乐 体育 资讯 电影 电视剧   片花 综艺 网络电影 脱口秀   动漫 生活 儿童 母婴 教育 健康 音乐 搞笑 时尚 原创 旅游 拍客 财经 军事 科技
d360022.txt 影视 对不起   您访问的页面不存在 首页 电视剧   电视剧首页   热门古装 超燃警匪 爆笑喜剧 青春偶像 综艺 综艺首页 真人秀   访谈 情感 电影 电影首页 搞笑大片 火爆动作 甜蜜热恋 恐怖鬼片
d361706.txt 您访问的页面不存在 射击装扮双人休闲儿童过关益智 更多 更多 首页 动作 测试 冒险 女生 棋牌 敏捷 体育 策略 综合 模拟经营 笑话漫画动画片手机游戏网页游戏   游戏推荐 完美漂移 迷你世界 星途少女
d366297.txt 404 非常抱歉 您访问的页面不存在 PP视频一起玩出精彩 原聚力视频   PPTV聚力 电影 首页 电视剧   电影 体育 综艺 动漫 频道 电视剧   偶像 爱情 军旅 历史 武侠 神话 电   影   动作 喜剧 爱情 恐怖
d366290.txt 404 -   非常抱歉，您访问的页面不存在 -   PP视频一起玩出精彩 -   原聚力视频、PPTV聚力   电影 首页 |   电视剧   电影 体育 综艺 动漫 频道 电视剧   偶像 爱情 军旅 历史 武侠 神话 电   影
d364377.txt 非常抱歉 您访问的页面不存在 视频一起玩出精彩   原聚力视频   聚力 电影 首页 电视剧   电影 体育 综艺 动漫 频道 电视剧   偶像 爱情 军旅 历史 武侠 神话 电   影   动作 喜剧 爱情 恐怖 犯罪 剧情
d363576.txt 404错误  你访问的页面丢失了 导航 更多频道内容在这里查看知道了 娱乐 体育 资讯 电影 电视剧   片花 综艺 网络电影 脱口秀   动漫 生活 儿童 母婴 教育 健康 音乐 搞笑 时尚 原创 旅游 拍客 财经 军事
d369481.txt 错误 你访问的页面丢失了 导航 更多频道内容在这里查看知道了 娱乐 体育 资讯 电影 电视剧   片花 综艺 网络电影 脱口秀   动漫 生活 儿童 母婴 教育 健康 音乐 搞笑 时尚 原创 旅游 拍客 财经 军事 科技
d360411.txt 未找到你访问的视频 中国第一资讯视频网站   热点资讯 民生资讯在线视频观看   首页 导航 健康 游戏 影视 时尚 专题 广场舞   儿童舞蹈 公益 瑜伽 原创 生活 美女 汽车 旅游 搜视频   视频 专辑 关闭
d360383.txt 游戏网   手机版   街机游戏TOP10   街机游戏TOP100   街机游戏TOP500   单机游戏TOP10   单机游戏TOP100   单机游戏TOP500   家机排行榜   掌机排行榜   模拟器排行榜  抱歉 您访
d360228.txt 51EDU精品学习网   欢迎来到精品学习网 登陆 免费注册  站内搜索  触屏版   最近更新  专题 很抱歉，您访问的页面不存在..... 提示：您可能输错了网址或者网页已删除，返回  网站首页或通过搜索重新查
d360036.txt 您访问的页面不存在 -   56.com  56.com导航>频道分类首页电视剧电影综艺56出品高校粤语搞笑音乐人生活游戏时尚美女科技教育汽车少儿母婴拍客用户服务自媒体分成应  用APP登录注册>上传上传视频>
d361720.txt 您访问的商铺不存在 供应 求购 公司 资讯 论坛 错误 没有找到您想要进入的页面   抱歉 您访问的页面不存在 您可以返回   上一页   或是 首页 相关搜索词   为您推荐热门产品   相关资讯列表   关于
d369037.txt 错误 你访问的页面丢失了 导航 更多频道内容在这里查看知道了 娱乐 体育 资讯 电影 电视剧   片花 综艺 网络电影 脱口秀   动漫 生活 儿童 母婴 教育 健康 音乐 搞笑 时尚 原创 旅游 拍客 财经 军事 科技
d367938.txt 您在万维商机网   访问的页面不存在或者已被删除 您访问的页面不存在或者已被删除   可以尝试如下方法找到你想要的 去万维商机网首页   看看 那或许有你想要的   点击浏览器回退按钮 看看上一页还有
d956060.txt 服务器错误：  404 页面未找到   |   首页 搜索 登录 首页 快讯 观察 电商号   关注订阅  打开微信“扫一扫”   登录 注册 微信登录 快讯 观察 电商号   关注订阅 快讯 观察 零售 物流 人物 B2B 支付
d955992.txt PC下载网—对不起，您访问的页面不存在   -PC下载网   小编推荐 发布软件 最近更新 全站导航  软件 电脑软件 安卓软件 苹果软件 游戏 安卓游戏 苹果游戏手游电脑版电脑游戏   资讯 资讯教程 软件
d771870.txt 热门检索：财政收支积极财政政策减税降费   首页 职能机构 新闻报道 信息公开 政务服务 交流互动 专题专栏 温馨提示：您访问的页面不存在或已删除  网站地图 联系我们 主办单位：中华人民共和国
d776425.txt 页面没有找到_搜铺网   抱歉！您访问的页面不存在！我们会尽快查找，提供您所需要的页面，请返回等待信息！   您不返回吗？  您确定不返回吗？   您真的确定现在不用返回吗？   好吧，随便您要不要
d774770.txt 抱歉 ，您访问的页面不存在   —   懂球帝   —   专业权威的足球网站|足球新闻|足球资讯|足球直播   51La   首页 比赛 数据 APP 懂球号   主播后台 登录 广告合作 抱歉 ，您访问的页面不存在……
d710516.txt 页面没有找到_搜铺网   抱歉！您访问的页面不存在！我们会尽快查找，提供您所需要的页面，请返回等待信息！   您不返回吗？  您确定不返回吗？   您真的确定现在不用返回吗？   好吧，随便您要不要
```

**Figure 3: Examples of Filtered Long Documents**

```
d397348.txt 您访问的页面不存在 抱歉 您访问的页面不存在 秒后为您跳转到精彩推荐  视频可能已被删除   网页地址可能有误   您也可以向我们反馈遇到的问题
d392701.txt 搜狗搜索
d390054.txt <unk>
d390987.txt 404 Not Found   404 Not Found   nginx
d393081.txt 您访问的页面不存在 抱歉 您访问的页面不存在 秒后为您跳转到精彩推荐  视频可能已被删除   网页地址可能有误   您也可以向我们反馈遇到的问题
d393050.txt <unk>
d391705.txt Not Found   Not Found   HTTP    Error   404 The requested   resource    is  not found
d397288.txt 404 Not Found   404 Not Found   nginx
d393843.txt Sorry   Page    Not Found
d391364.txt 搜狗搜索
d396656.txt 页面找不到    潇湘书院  找不到页面    对不起    您访问的页面不存在或已被删除 返回之前访问的页面  或者 访问 潇湘书院首页
d394562.txt <unk>
d397477.txt 403 Forbidden   403 Forbidden   nginx
d392329.txt 您访问的页面不存在 抱歉 您访问的页面不存在 秒后为您跳转到精彩推荐  视频可能已被删除   网页地址可能有误   您也可以向我们反馈遇到的问题
d393569.txt <unk>
```

**Figure 4: Examples of Meaningless Candidate Documents**

$$score(query, doc) = MLP(CLS[BERT(Input)])$$

where CLS is the BERT's [cls] vector and MLP is a Multilayer Perceptron that projects the CLS vector to a score. We use the Localized Contrastive Estimation (LCE)[6] loss function to optimize our model, which is defined as:

$$Loss := -\log \frac{exp(score(q, d^{+}))}{\sum_{d \in S} exp(score(q, d))}$$

where S is the set of all candidate documents, q is the query and $d^{+}$ is the clicked document.

We use the same Transformer encoder architecture as BERT[4]. The hidden size is 768 and the number of self-attention heads is 12. We use the Adam optimizer with a learning rate of 1e-5 and a warm-up ratio of 0.1. The batch size is set as 32 and the maximum length of the input sequence is set as 256. We save the checkpoints every 1000 steps and choose the model with the highest NDCG@3 score as the final model.

*2.4.2 Fine-tune with Session Information.* In a search session, a single query may not satisfy a user's information need, thus the user submits more queries. During the search process, the user may click some documents provided by the search engine, those documents can also reflect the user's information needs. Therefore, for a query in a search session, we concatenate the previously clicked documents' titles and the previous queries to formulate an assembled query which we call Assembled Session Query(ASQ). The ASQ incorporates information from the entire search session.

We fine-tune the BERT model with ASQ which is defined as follows:

$$cq := CurrentQuery,$$

$$pq := PreviousQueries,$$

$$tcd := ClickedDocumentsTitle$$

$$ASQ = [CLS]cq[SEP]pq[SEP]tcd$$

$$Input := [CLS]ASQ[SEP]doc[SEP]$$

$$score(query, doc) = MLP(CLS[BERT(Input)])$$

The loss function and other settings of the fine-tuning stage are the same as Section 2.4.1.

*2.4.3 Experimental Results and Analysis.* The result of preliminary evaluation shows that the utilization of session information can improve the performance of ranking models to a small degree. We can conclude that fine-tuning with user behavior information can improve the document ranking model's performance because the Assembled Session Query(ASQ) contains more information of user's search intention compared to a single query. However, it can only improve the performance to a small degree because the user's search intention may change in a search session. In this situation, previous clicked documents' titles may not reflect the user's current search intention.

## 2.5 Learning-to-rank model

To assemble all those scores and features introduced in the previous section, we feed all of them to the LambdaMART model with the default parameters and the training metric NDCG@10. The features are shown in Table 3. This run achieves the best performance among all participants in the preliminary evaluation. The good performance shows that the scores of fine-tuned pre-trained language model can improve the performance of the LambdaMART model.

```
SessionID   11
---------------------------
4399赛尔号  q20 1427845508.16
1  http://www.4399.com/flash/seer.htm  d209    赛尔号_4399赛尔号游戏在线玩_赛尔号精灵大全_赛尔号攻略   1   1427845518.218
2  http://www.4399.com/    d210    小游戏,4399小游戏,小游戏大全,双人小游戏大全 0   -1
3  http://news.4399.com/gonglue/seerye.html    d211    <unk>   0   -1
4  http://news.4399.com/seer2/ d212    约瑟传说_赛尔号2已改名约瑟传说_赛尔号2精灵大全_4399赛尔号2游戏  0   -1
5  http://wenwen.sogou.com/s/?sp=S4399%E8%B5%9B%E5%B0%94%E5%8F%B7  d213    搜狗搜索    0   -1
6  http://xiaoyouxi.2345.com/flash/12097.htm    d214    【4399赛尔号】7k7k赛尔号-赛尔号精灵大全-2345小游戏大全  0   -1
7  http://weixin.qq.com/   d5  微信，是一个生活方式    0   -1
8  http://baike.sogou.com/v9833876.htm d215    4399赛尔号  0   -1
9  http://seer.61.com/ d216    赛尔号  0   -1
10 http://saier.wandodo.com/    d217    <unk>   0   -1
---------------------------
4399赛尔号  q20 1427846953.13
1  http://www.4399.com/flash/seer.htm  d209    赛尔号_4399赛尔号游戏在线玩_赛尔号精灵大全_赛尔号攻略   1   1427846955.343
2  http://www.4399.com/    d210    小游戏,4399小游戏,小游戏大全,双人小游戏大全 0   -1
3  http://news.4399.com/gonglue/seerye.html    d211    <unk>   0   -1
4  http://news.4399.com/seer2/ d212    约瑟传说_赛尔号2已改名约瑟传说_赛尔号2精灵大全_4399赛尔号2游戏  0   -1
5  http://wenwen.sogou.com/s/?sp=S4399%E8%B5%9B%E5%B0%94%E5%8F%B7  d213    搜狗搜索    0   -1
6  http://xiaoyouxi.2345.com/flash/12097.htm    d214    【4399赛尔号】7k7k赛尔号-赛尔号精灵大全-2345小游戏大全  0   -1
7  http://weixin.qq.com/   d5  微信，是一个生活方式    0   -1
8  http://baike.sogou.com/v9833876.htm d215    4399赛尔号  0   -1
9  http://seer.61.com/ d216    赛尔号  0   -1
10 http://saier.wandodo.com/    d217    <unk>   0   -1
```

**Figure 5: The Format of Training Data**

**Table 3: Features of PMTM Model**

| | Features |
|---|---|
| 1 | Document Length |
| 2 | Query Length |
| 3 | Document ID |
| 4 | Score of BERT finetuned with ad-hoc data |
| 5 | Score of BERT finetuned with session data |
| 6 | Score of BM25 |
| 7 | Score of TF-IDF |
| 8 | Score of F1-EXP |
| 9 | Rank sorted by BM25 |
| 10 | Rank sorted by TF-IDF |
| 11 | Rank sorted by F1-EXP |

## 3 POSS SUBTASK

In the POSS subtask, we use the LambdaMART model with the same features in Section 2.4.3 which is shown in Table 3. The performance of our run ranked first place in the preliminary evaluation and second place in the final evaluation.

## 4 CONCLUSIONS

In the NTCIR-16 Session Search (SS) Task, we participated in both FOSS and POSS subtasks. We tried traditional methods as well as pre-trained language model. During the fine-tuning stage, we compare the performance of utilizing the session information and ad-hoc information. The result shows that the utilization of session information can improve the performance of ranking models to a small degree. In the final evaluation, the assembled traditional methods beat complicated pre-trained language model which shows the effectiveness of traditional methods on document ranking tasks.